\documentclass[aps,twocolumn,showpacs]{revtex4}
\usepackage{psfig}

\def\ep{\epsilon}
\def\order#1{{\cal O}\left(#1\right)}
\newcommand{\ba}{\begin{eqnarray}}
\newcommand{\ea}{\end{eqnarray}}
\newcommand{\be}{\begin{equation}}
\newcommand{\ee}{\end{equation}}
\def\eq#1{(\ref{#1})}

\begin{document}

\title{
%
%
\[ \vspace{-2cm} \]
\noindent\hfill\hbox{\rm  } \vskip 1pt
\noindent\hfill\hbox{\rm Alberta Thy 15-01} \vskip 1pt
\noindent\hfill\hbox{\rm SLAC-PUB-9013} \vskip 1pt
\noindent\hfill\hbox{\rm hep-ph/0110028} \vskip 10pt
Threshold expansion for heavy-light systems \\
and flavor off-diagonal current-current correlators
}

\author{Andrzej Czarnecki}
\affiliation{
Department of Physics, University of Alberta\\
Edmonton, AB\ \  T6G 2J1, Canada\\
E-mail:  czar@phys.ualberta.ca}

\author{Kirill Melnikov}
\affiliation{
Stanford Linear Accelerator Center\\
Stanford University, Stanford, CA 94309\\
E-mail: melnikov@slac.stanford.edu}

\begin{abstract}
An expansion scheme is developed for Feynman diagrams describing the
production of one massive and one massless particle near the
threshold.  As an example application, we compute the correlators of
heavy-light quark currents, $\bar b \gamma_\mu u$, $\bar b \gamma_5
u$, through ${\cal O}(\alpha_s^2)$.
\end{abstract}


\maketitle

Processes mediated by $W$-bosons often involve production or
annihilation of two fermions with very different masses.  Examples of
present interest include the $B$-meson decay constant $f_B$ and the single
top-quark production.  The basic ingredient in the analysis of such
processes are correlators of heavy-light currents.   For example,
$f_B$ can be determined by relating such correlators to  
the measured spectrum of $B$ mesons, with help of the QCD sum rules
\cite{Khodjamirian:2001bj}.

In another important application one can use such correlators,
computed perturbatively in the continuum, as an input for lattice
calculations, in particular for the matching of the lattice and
continuum currents.  The perturbative production cross section
computed close to the threshold (related to the imaginary part of the
current correlators) is a convenient physical observable to 
perform the matching \cite{lepage}.

\begin{figure}[htb]
\hspace*{-38mm}
\begin{minipage}{16.cm}
\begin{tabular}{ccc}
\psfig{figure=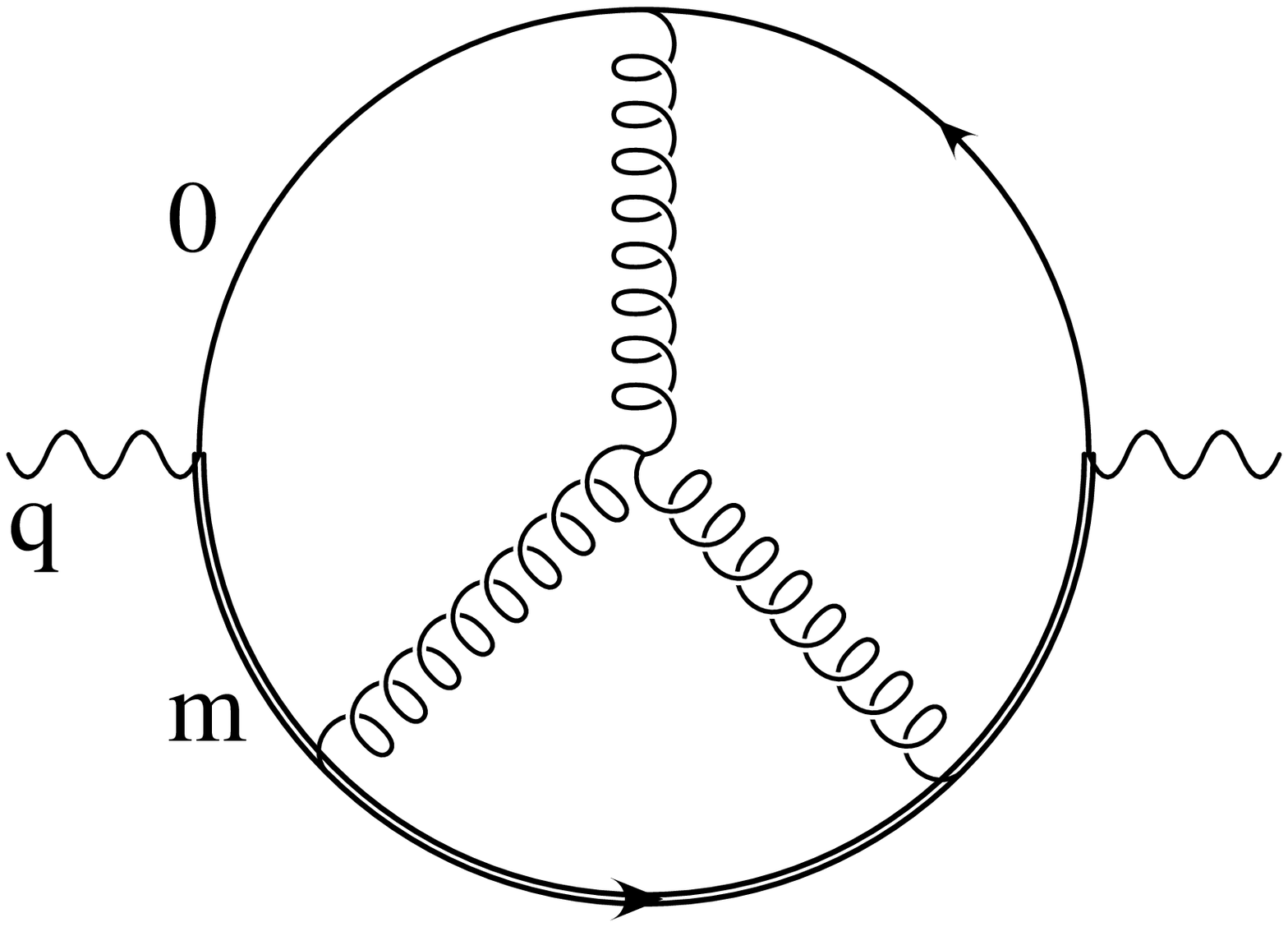,width=27mm}
&\hspace*{0mm}
\psfig{figure=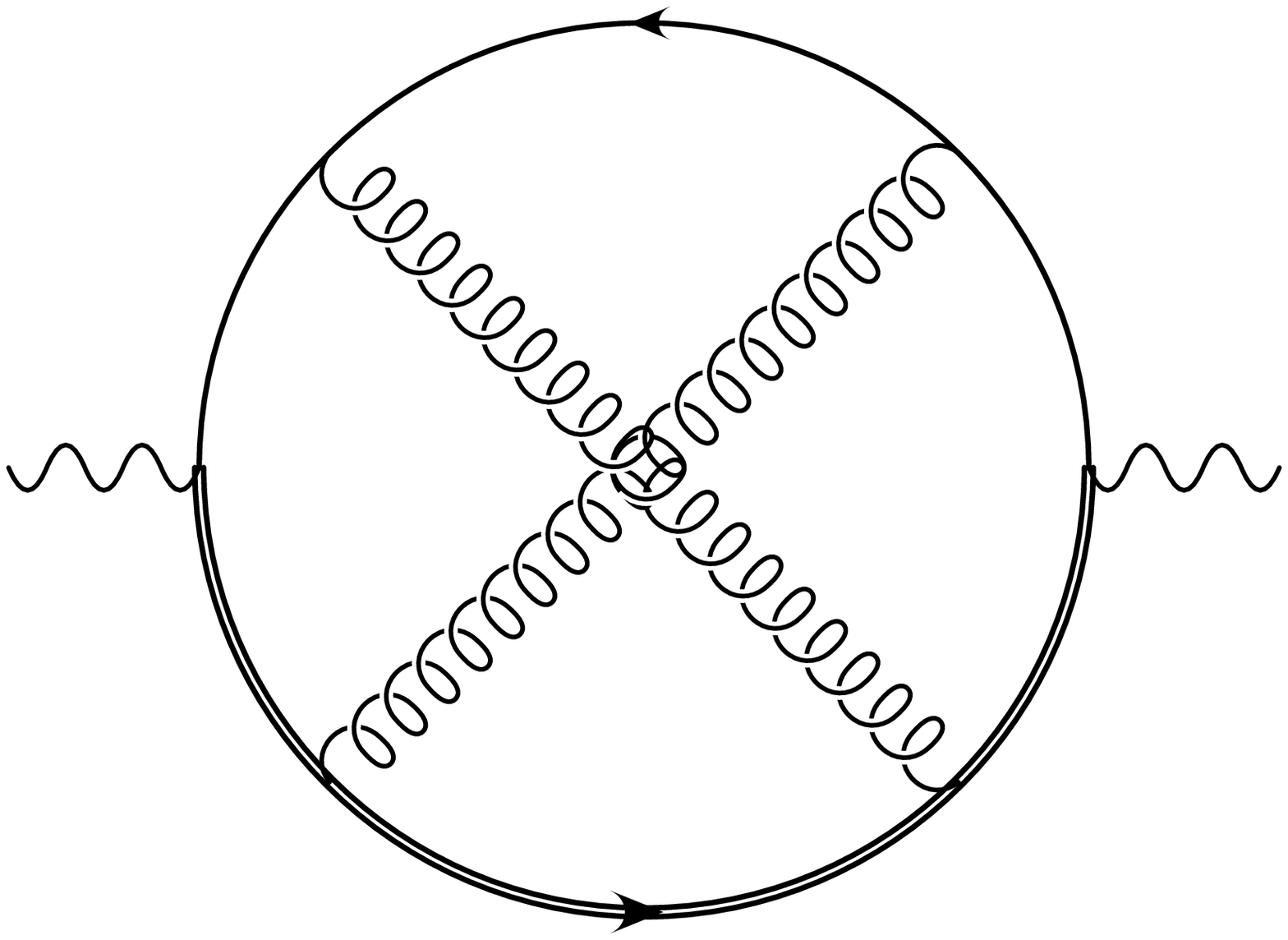,width=27mm}
&\hspace*{0mm}
\psfig{figure=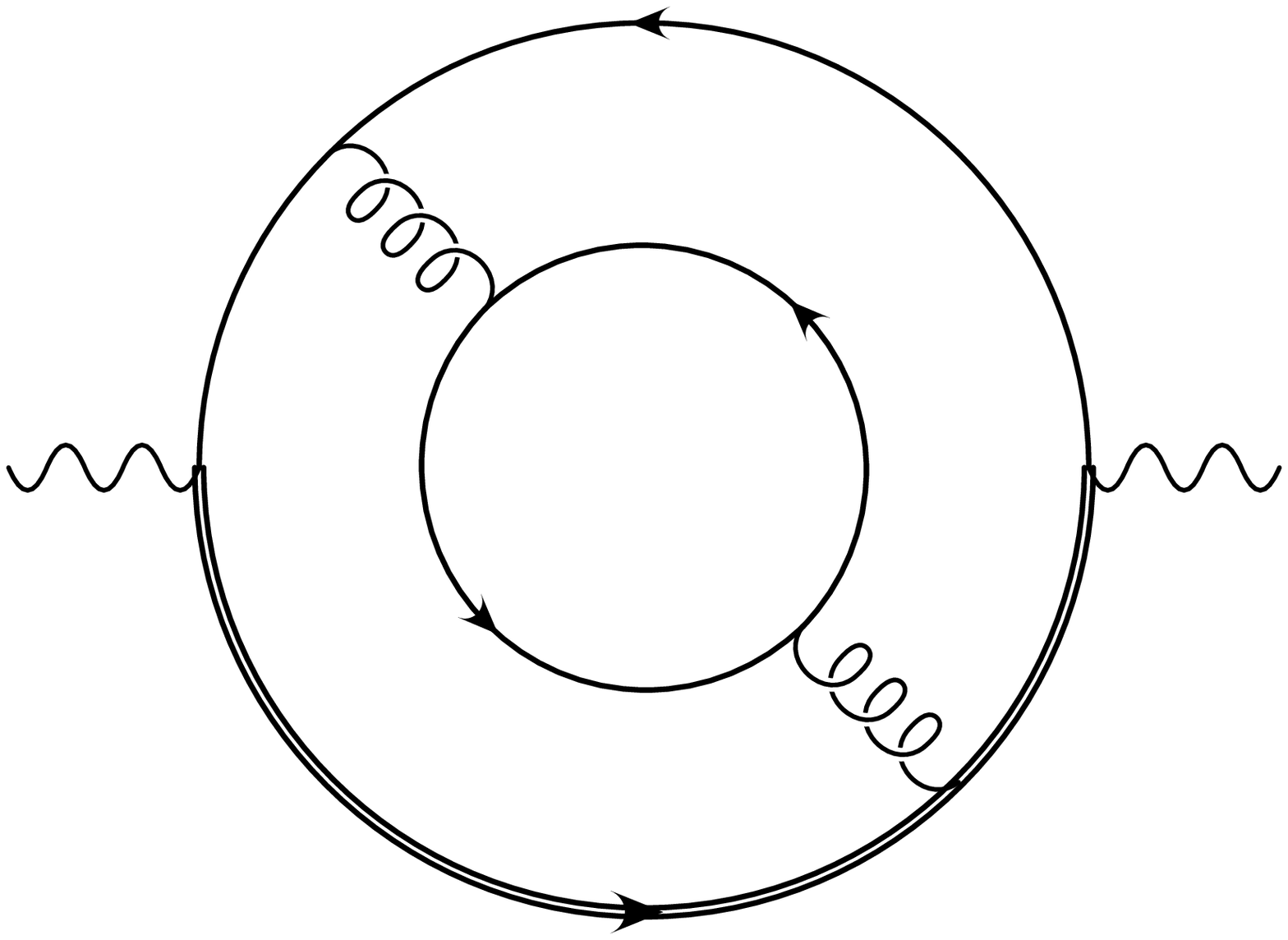,width=27mm}
\end{tabular}
\end{minipage}
\caption{Examples of diagrams contributing to the current correlators
at $\order{\alpha_s^2}$.}
\label{fig:cor}
\end{figure}

Even if the light fermion mass is neglected, a correlator still
involves two mass scales (see Fig.~\ref{fig:cor}), 
the invariant mass of the pair $\sqrt{q^2}$
and the mass of the heavy fermion $m$, which hampers the evaluation of
the higher-order quantum effects.  It is helpful to develop a
computational scheme which allows to expand the Feynman diagrams
around the threshold, $q^2=m^2$.  In the past, various expansion
schemes were constructed for other kinematical situations
\cite{Chetyrkin91,Tkachev:1994gz,Smirnov:1995tg,%
Smirnov:1997gx,CzSmir,Beneke:1998zp,Czarnecki:2000fv}.
They facilitated many studies of
higher-order radiative corrections to a variety of processes of
experimental importance, such as the $Z$-boson decays, heavy quark
production and decays, electromagnetic properties of particles, etc.
In the present paper we present an expansion scheme for the case
depicted in Fig.~\ref{fig:cor}, for $q^2\simeq m^2$, and apply it to compute
$\order{\alpha_s^2}$ corrections to the imaginary part of the
heavy-light correlators of (axial)vector as well as (pseudo)scalar
currents.  The real part can be obtained, if need arises, along the same
lines.

Recently, numerical values of those corrections were estimated
\cite{Chetyrkin:2000mq,Chetyrkin:2001je} using Pad\'e approximants to
describe the 
correlator as a function of $q^2$.  The approximants were obtained
from several terms in the expansion of the correlator for large ($q^2
\gg m^2$) and small ($q^2 \ll m^2$) values of the total energy.
This approach is accurate for $q^2$ sufficiently far from the
production threshold $q^2=m^2$. However, for practical applications
related to  heavy quark physics, one needs the
correlators quite close to the threshold where, for kinematic reasons,
their imaginary part vanishes as $(q^2-m^2)^2$.  Because of this
suppression, $\order{\alpha_s^2}$ terms obtained from the Pad\'e
approximants are rather uncertain near the
threshold, with an error estimated to be about
$30\%$ \cite{Chetyrkin:2001je}.

The approach presented in this Letter enables an analytic calculation
of those terms.  We introduce the parameter $\delta = 1-m^2/q^2$ which
is small in the kinematic region close to threshold.  We will show
that an expansion in $\delta$ can be constructed applying the
reasoning familiar from HQET directly to the Feynman diagrams.  The
only new ingredient necessary for a calculation of the ${\cal
O}(\alpha_s^2)$ corrections to the correlators are certain three-loop
HQET diagrams whose calculation we will describe.

We consider the following correlator,
\ba
\lefteqn{ \left (-q^2g_{\mu\nu}+q_\mu q_\nu \right ) 
\Pi^{v}(q^2) +q_\mu q_\nu \Pi_L^{v}(q^2) }
\nonumber \\
&&= i \int {\rm d}xe^{iqx} \langle 0| Tj_\mu^v(x)j_\nu^v(0)|0\rangle ,
\ea
with $j_\mu^v = \bar \psi_1 \gamma_\mu \psi_2$ and $\psi_{1,2}$
denoting the Dirac spinors for the massless and massive quarks,
respectively. This correlator is an analytic function of $q^2$ with a
cut starting at $q^2=m^2$, corresponding to $\delta=0$.  Since
$\delta$ is proportional to the phase space available for the
production of heavy and light quarks, for small $\delta$ the heavy
quark is non-relativistic and always close to its mass shell.  On the
contrary, the massless quark is always ultrarelativistic, but its
energy is small if $\delta$ is small.  The HQET is designed to study
exactly this kind of situation and the calculation of the relevant
Feynman diagrams can be simplified if one follows its pattern.

We have to consider two different scales of the momentum $k$
flowing along a given line in a Feynman diagram: hard $k \sim m$
or soft $k \sim m \delta$.  Consider the heavy quark propagator
$\sim 1/(k^2 + 2qk + q^2 \delta)$. If $k$ is hard,  the propagator
can be expanded in a Taylor series in $\delta$ yielding the on-shell
heavy quark propagator $ \sim 1/(k^2 + 2qk)$. On the other hand, if
$k$ is soft, the propagator can be expanded in $k^2$  resulting
in the static heavy quark propagator $\sim 1/(2qk + q^2
\delta)$, familiar from HQET.  

For each diagram, one has to consider all possible momentum routings
and find all contributing subgraphs. Among them, there are two which
can be easily described. First there is the situation when all lines
are soft so that all heavy quark propagators become static and the
diagram becomes what is usually referred to as a HQET matrix
element. In our case, these subgraphs require three-loop calculations
in HQET and we will explain below how we solve this problem.

The second type of subgraphs arises in the situation when all momenta
are hard.  In this case all heavy quark propagators are Taylor
expanded in $\delta$; the resulting Feynman diagrams are of the
on-shell three-loop propagator type, studied in
\cite{Laporta:1993pa,Melnikov:2000zc}, and their evaluation is
possible. However, since these contributions are polynomials in
$\delta$, they do not contribute to the imaginary part of the
correlator and we do not consider them here.

In between the two extreme cases discussed above, there are situations
where, in a given diagram, some of the lines are soft and some are
hard. Using the HQET language, these correspond to the HQET matrix
elements with insertions of higher dimensional operators of the HQET
Lagrangian or to the HQET matrix elements computed with leading order
operators (HQET currents) corrected for the higher order Wilson
coefficients. Since the Wilson coefficients are computed at the hard
scale, such contributions factorize into products of simple subgraphs
and can be easily computed.

Therefore, the main challenge are the three-loop HQET diagrams. There
are two ways to compute them and we have taken both to have a cross
check. Recently, the three-loop HQET diagrams have been analyzed in
\cite{Grozin:2000jv,Grozin:2001fw} and a computer algebra program has
been published, capable of computing all three-loop HQET propagator
type diagrams.  We have used that software to calculate the required
HQET matrix elements.  For the purpose of the cross check, we have
written (in FORM \cite{form3}), in a completely independent way, a
similar program solving the three-loop recurrence relations for HQET
propagator-type integrals, restricting ourselves to topologies needed
for the current calculation.

In both approaches, every Feynman diagram is expressed in terms of a
few master integrals.  The majority of them is known
\cite{Grozin:2000jv,Grozin:2001fw,Beneke:1994sw}. However, one 
of the master integrals has not been evaluated to sufficient accuracy
in the existing literature and we have to compute it. It turns
out that one can use a trick  to this end.  The Euclidean 
integral we need is ($p^2 = -1$, $d=4-2\epsilon$):
\ba
I &=& \int 
\frac{{\rm d}^d k_1 {\rm d}^d k_2 {\rm d}^d k_3 }{k_1^2 k_3^2 (k_1 -
k_2)^2 (k_3 - k_2)^2}
\nonumber \\
&&  \times  \frac{1}{ (2pk_1 + 1) (2 pk_2 + 1)  (2 p k_3 + 1)}.
\ea
To compute it, consider a similar integral,
\ba
I_1 &=& \int 
\frac{{\rm d}^d k_1 {\rm d}^d k_2 {\rm d}^d k_3 }{k_1^2 k_2^2 k_3^2
(k_1 - k_2)^2 (k_3 - k_2)^2} 
\nonumber \\
&& \times  \frac{1}{ (2pk_1 + 1) (2 pk_2 + 1) (2 p k_3 + 1)},
\label{eq3}
\ea
related to $I$ by integration-by-parts \cite{che81} identities.  We 
perform  
a transformation $|k_i| \to 1/|k_i|$ for $i=1,2,3$ in Eq.~(\ref{eq3}).
The integral $I_1$ transforms to
\ba
I_1 &=& \int 
\frac{{\rm d}^d k_1 {\rm d}^d k_2 {\rm d}^d k_3 }{k_1^{2d-6}
k_2^{2d-8} k_3^{2d-6} (k_1 - k_2)^2 (k_3 - k_2)^2} 
\nonumber \\
&& \times \frac{1}{ (2pk_1 + k_1^2) (2 pk_2 + k_2^2) (2 p k_3 + k_3^2)}.
\ea
In the next step we notice that $I_1$ is finite in four dimensions, 
so that the limit $d \to 4$ can be taken; after that
$I_1$ becomes equal to one of the on-shell three-loop master integrals
computed in \cite{Melnikov:2000zc}. We therefore find:
\be
I_1 = C(\ep) \left [ 2\pi^2 \zeta_3 - 5 \zeta_5 \right ] +{\cal O}(\ep),
\ee 
with $C(\ep) = [\pi^{2-\ep}\Gamma(1+\ep)]^3$.
We now use recurrence relations to obtain $I$ from $I_1$ and derive:
\ba
I &=& C(\ep) \left [ -\frac{\pi^2}{18\ep^2}
  + \frac{1}{\ep} \left (-\frac{4\pi^2}{9}+\frac{1}{3}\zeta_3 \right )
\right. \nonumber \\
&& \left.  
+ \left ( \frac{8}{3}\zeta_3- \frac{26\pi^2}{9}-\frac{17\pi^4}{540} \right )
 + \ep \left (-\frac{4\pi^2}{9}\zeta_3 
\right. \right.
\nonumber \\
&& \left. \left.
+ \frac{52}{3}\zeta_3
 -\frac{160\pi^2}{9}- \frac{34 \pi^4}{135}- \frac{83}{3}\zeta_5 \right )
+{\cal O}(\ep^2)
\right ].
\ea
With this integral at hand, all the three-loop HQET master integrals
are available to sufficiently high power in their expansion in $\ep$ and 
we can proceed with the computation of the correlators.

\begin{figure}[htb]
\hspace*{-38mm}
\begin{minipage}{16.cm}
\begin{picture}(100,0)
\put (23,-30) {$\Delta^v_{\rm A}$}
\put (150,-30) {$\Delta^v_{\rm NA}$}
\put (23,-120) {$\Delta^v_{\rm L}$}
\put (150,-120) {$\Delta^v_{\rm H}$}
\end{picture}

\begin{tabular}{cc}
\psfig{figure=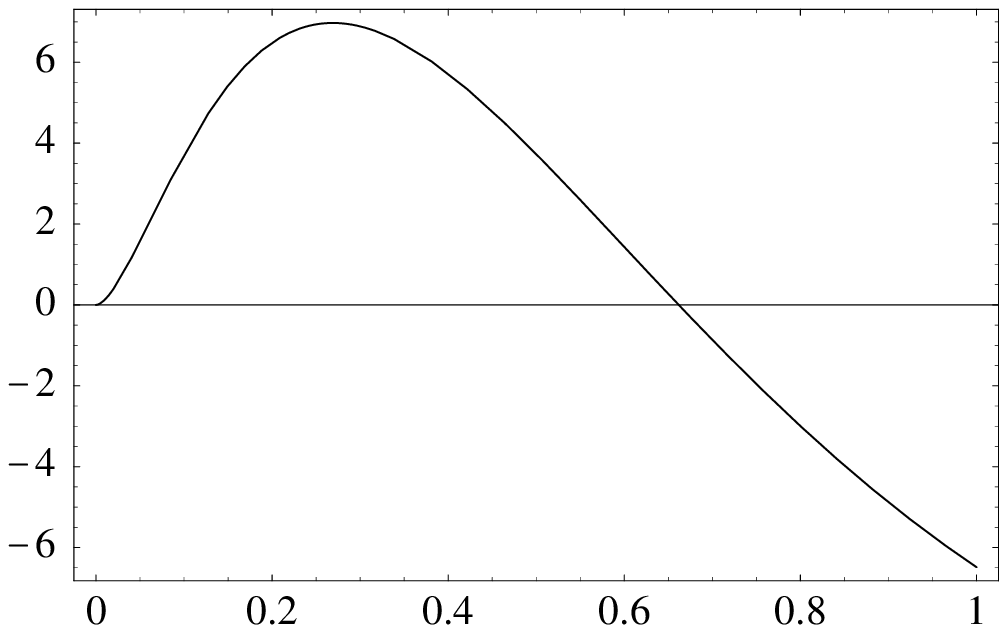,width=40mm}
&\hspace*{4mm}
\psfig{figure=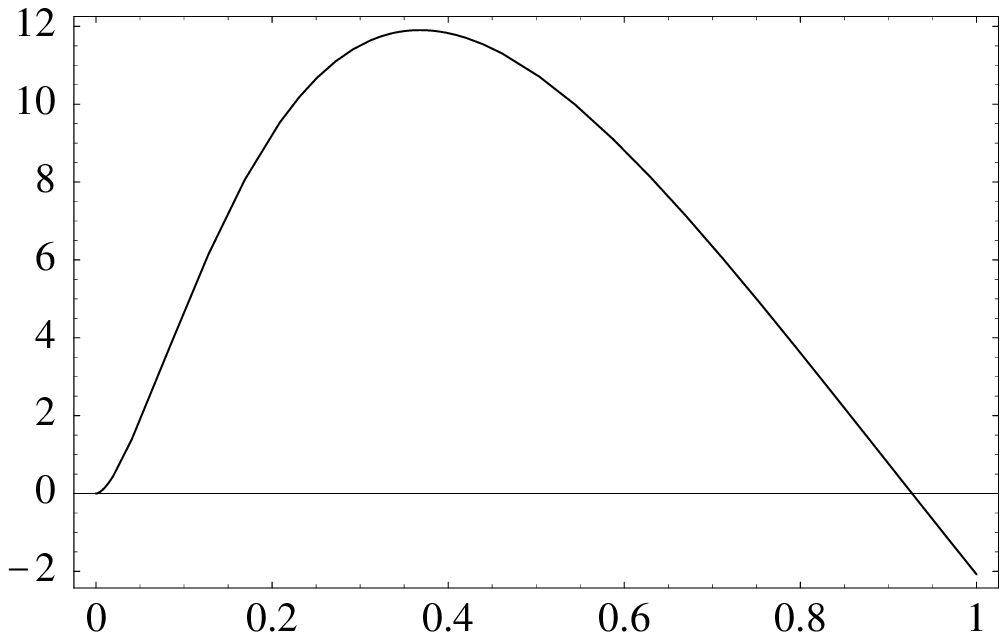,width=40mm}
\\[1mm]
\psfig{figure=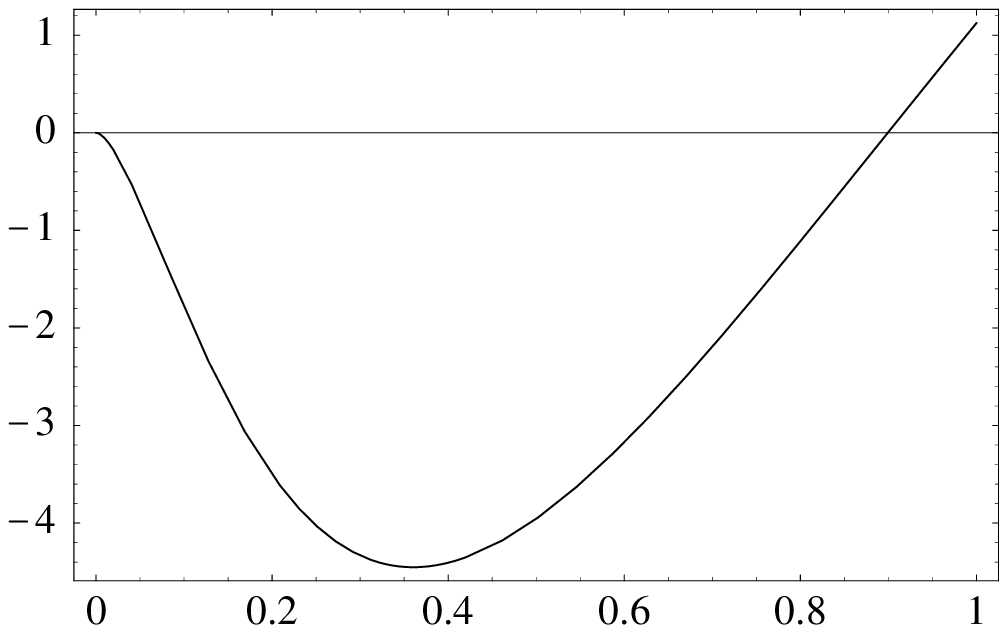,width=40mm} 
&\hspace*{4mm}
\psfig{figure=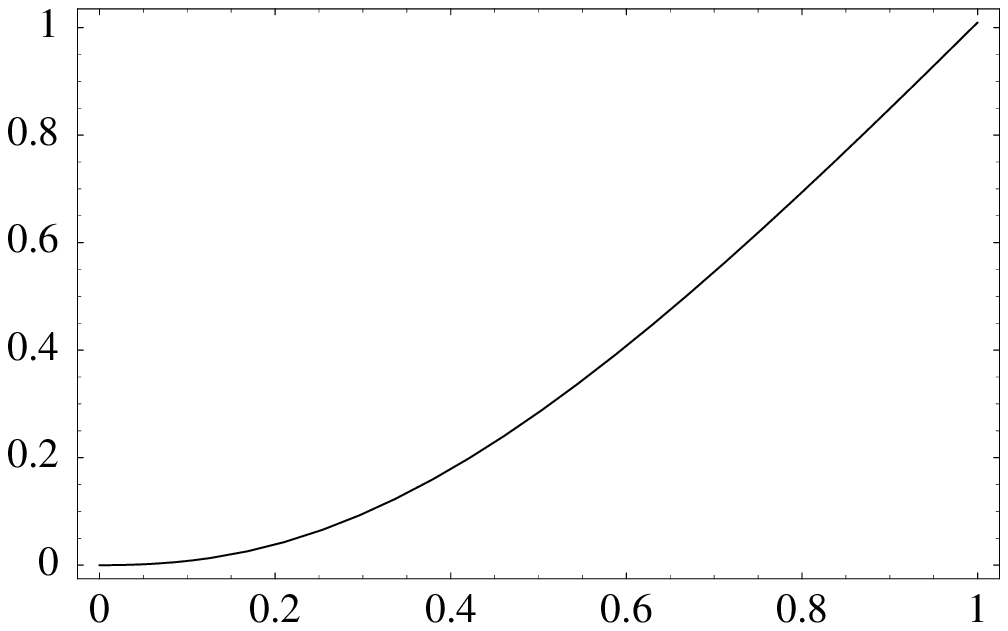,width=40mm}
\\[1mm]
\end{tabular}
\end{minipage}
\caption{$N_c \delta^2 \Delta^v_{\rm A,NA,L,H}$
plotted as a function of $v = \delta/(2-\delta)$.
} 
\label{fig:vector}
\end{figure}

We introduce the dimensionless quantity
\be
R^{v}(q^2) \equiv{\Gamma(W^\star \to b\bar u) \over |V_{bu}|^2
 \Gamma(W^\star \to e\bar \nu)}
  = 12 \pi {\rm Im} \left [ \Pi^{v}(q^2+i\delta) \right ],
\ee
and expand it in a series in the strong coupling constant (we use the
$\overline{\rm MS}$ scheme and denote $a \equiv  \alpha_s(M)/\pi$)
\be
R^{v} \simeq N_c \left ( R_0^v 
+ C_F a R_1^v 
+ C_F a^2 R_2^v  \right ),
\ee
We find (an exact formula for $R^{(1),v}\equiv N_c R_1^{v}$ can be found 
in \cite{Chetyrkin:2001je}, eq.~(9)):
\ba
R_0^v &=& \frac{\delta^2 (3 - \delta)}{2}
\nonumber \\
R_1^v &=&\frac{3\delta^2}{4} \left [ \left ( \frac{9}{2} - 3\ln \delta 
+ \frac{2}{3} \pi^2 \right ) 
\right. \nonumber \\
&& \left. + \delta \left (\frac{-139}{18} 
+ \frac{11}{3} \ln \delta -\frac{2}{9} \pi^2 \right ) \right ]
+ {\cal O}(\delta^4),
\nonumber \\
 R_2^v &=& \delta^2 \left ( 
C_F \Delta_A^v + C_A \Delta_{NA}^v + T_R N_L \Delta_L^v 
+ T_R \Delta_H^v 
\right ). 
\ea
For the individual color structures we obtain:
\ba
\lefteqn{
 \Delta_A^v = 
 \frac{523}{64} - \frac{39}{8}\zeta_3 + \frac{7\pi^2}{4} \ln 2 
  + \frac{5\pi^2}{12} + \frac{\pi^4}{60}}
\nonumber \\
&& 
- \left( \frac{5\pi^2}{4} + \frac{147}{32} \right )L_\delta
     + \frac{27}{16} L_\delta^2 
\nonumber \\
&&+ \delta \left [ 
 - \frac{10079}{576} + \frac{39}{8}\zeta_3 - \frac{19\pi^2}{12}\ln 2
     - \frac{575 \pi^2}{432} - \frac{\pi^4}{180}
\right. 
\nonumber \\
&& \left.      
+ \left ( \frac{53\pi^2}{36}
          + \frac{1339}{96} \right )L_\delta
- \frac{57}{16}L_\delta^2 \right ] 
+ \delta^2 \left [ 
\frac{1819}{144}  
\right . \nonumber \\
&& \left. 
- \frac{5\pi^2}{24}\ln 2 
+ \frac{421 \pi^2}{1728}
  + \left ( \frac{\pi^2}{72} - \frac{395}{48} \right ) L_\delta 
    + \frac{17}{16} L_\delta^2 \right ],
\nonumber \\ 
\lefteqn{ \Delta_{NA}^v = 
\frac{1103}{64} - \frac{129}{16}\zeta_3 - \frac{7\pi^2}{8}\ln 2 + 
         \frac{427}{288}\pi^2 - \frac{\pi^4}{60} 
}
\nonumber \\
&& 
- \left ( \frac{19\pi^2}{24} + \frac{423}{32} \right )L_\delta 
+ \frac{33}{16} L_\delta^2  
\nonumber \\
&&
+ \delta \left [ 
- \frac{42655}{1728}  
 + \frac{33}{16}\zeta_3 
 + \frac{19\pi^2}{24}\ln 2 - \frac{\pi^2}{9} + \frac{\pi^4}{180}
\right.
\nonumber \\
&& \left.
 - \left (\frac{\pi^2}{24}
          - \frac{4685}{288} \right ) L_\delta 
- \frac{115}{48}L_\delta^2  \right ]
+ \delta^2 \left [ 
\frac{7931}{3456} - \frac{\zeta_3}{16}
\right.
\nonumber \\
&& \left. 
 + \frac{5}{48}\pi^2\ln 2 - \frac{115}{1152}\pi^2
 - \left ( \frac{\pi^2}{48} - \frac{305}{576} \right ) L_\delta 
 - \frac{17}{48}L_\delta^2  
\right ],
\nonumber \\ 
\lefteqn{ \Delta_{L}^v =
  - \frac{117}{16} + 3\zeta_3 - \frac{7\pi^2}{36} 
+ \left ( \frac{\pi^2}{3} + \frac{39}{8} \right ) L_\delta 
 - \frac{3}{4} L_\delta^2
}
\nonumber \\
&& +\delta
\left[ \frac{3671}{432}  - \zeta_3 - \frac{5\pi^2}{108} 
- \left (\frac{\pi^2}{9} + \frac{397}{72} \right )L_\delta
 + \frac{11}{12} L_\delta^2  
\right]
\nonumber \\
&& +\delta^2 \left (  - \frac{721}{864} + \frac{1}{18}L_\delta 
+ \frac{1}{12}L_\delta^2 \right ),
\nonumber \\
\lefteqn{ \Delta_{H}^v = 
    {133\over 32} - {5\pi^2 \over 12}
       + \delta  \left(  - {1997\over 864} + {\pi^2\over 4}\right)
}
\nonumber \\
&& 
       + \delta^2  \left( {2473\over 10800} - {1\over 20}L_\delta 
 - {\pi^2\over72} \right).
\label{eq:v}
\ea
We denoted $L_{\delta} = \ln \delta$ and the zeta function
$\zeta_3\simeq 1.202$. We do not display higher order terms in
the  expansion in $\delta$, but they can be easily obtained as well.

For completeness, we also give the results for ${\cal O}(\alpha_s^2)$ 
correction to the correlator of two scalar currents. Let us define
\ba
q^2\Pi^{s}(q^2)
= i \int {\rm d}xe^{iqx} \langle 0| Tj^s(x)j^s(0)|0\rangle,
\ea
where $j_s = Z_{\rm m} \bar \psi_1 \psi_2$ and, again, $\psi_{1,2}$ 
denote the Dirac spinors for the massless and massive quarks 
and $Z_{\rm m}$ is the on-shell  mass renormalization constant for 
the massive quark. We define
\ba
R^{s}(q^2) &=& 8 \pi {\rm Im} \left [ \Pi^{s}(q^2+i\delta) \right ],
\nonumber \\
 &\simeq & N_c \left ( R_0^s 
+ C_F a R_1^s 
+ C_F a^2 R_2^s  \right ).
\ea
We then find (for an exact formula for $R_1^{s}$ see
\cite{Chetyrkin:2001je}) 
\ba
R_0^s &=& \delta^2,
\nonumber \\
R_1^s &=&\delta^2 \left [ \left ( \frac{13}{4} - \frac{3}{2}L_\delta 
+ \frac{1}{3} \pi^2 \right ) 
-3 \delta \right ]
+ {\cal O}(\delta^4),
\nonumber \\
 R_2^s &=& \delta^2 \left ( 
C_F \Delta_A^s + C_A \Delta_{NA}^s + T_R N_L \Delta_L^s 
 + T_R \Delta_H^s 
\right ). 
\nonumber \\
\ea
For the individual color structures we obtain:
\ba
\lefteqn{\Delta_{A}^s = 
 {229 \over 32} 
 + {\pi^2 \over 2} \ln 2 + {3 \over 2} \pi^2
 + {\pi^4 \over 90}  - {9 \over 4} \zeta_3 }
\nonumber \\ 
&&
 - \left({5 \over 6} \pi^2 + {73 \over 16}\right)L_\delta
 + {9 \over 8} L_\delta^2
\nonumber \\ 
&& 
 + \delta   \left[  - {493 \over 48}
- {115 \over 72} \pi^2  - {3 \over 2} \zeta_3  
  +\left( 4-{\pi^2\over 3} \right) L_\delta
 \right] 
\nonumber \\
&&
 + \delta^2   \left[  - {4325 \over 864}  - {\pi^2 \over 4} \ln 2 -
{671 \over 864} \pi^2+ {1 \over 2} \zeta_3
  \right.
\nonumber \\ &&
\left.\qquad
+ \left({245\over 36} + {7 \over 36} \pi^2\right)L_\delta
          - {29 \over 24} L_\delta^2
 \right]
\nonumber \\
\lefteqn{\Delta_{NA}^s = 
         {4361 \over 288} - {\pi^2 \over 4} \ln 2 + {283 \over
432} \pi^2 - {\pi^4 \over 90}  - {47 \over 8}  
         \zeta_3 }
\nonumber \\
&& \qquad
 - \left({19 \over 36} \pi^2 + {141 \over
16}\right)L_\delta +  
         {11 \over 8} L_\delta^2
\nonumber \\
&&
 + \delta \left[ -{949\over 72} - {31 \over 432} \pi^2 - {\zeta_3 \over 4}
 + \left( {11 \over 3} - {\pi^2 \over 18} \right)L_\delta 
       - {1 \over 4} L_\delta^2
\right]
\nonumber \\
&&       + \delta^2 \left[  - {4651 \over 864} + {\pi^2 \over 8} \ln 2 
         + {179 \over 1728} \pi^2   -  
         {1 \over 4} \zeta_3
  \right.
\nonumber \\ &&
\left.\qquad
 + \left( {283 \over 96} - {7 \over 72}\pi^2 \right)L_\delta 
          - {5 \over 12} L_\delta^2
 \right]
\nonumber \\
\lefteqn{\Delta_{L}^s = 
 - {427 \over 72} - {7 \over 54}\pi^2 + 2 \zeta_3 
 + \left({2 \over 9} \pi^2
 + {13 \over 4}\right)L_\delta - {1 \over 2} L_\delta^2} 
\nonumber \\
&&       + \delta 
\left( {163 \over 36} + {2 \over 9} \pi^2 - {4 \over 3} L_\delta \right)
 + \delta^2 \left( {623 \over 432} - {19 \over 18}L_\delta +{1 \over 6} 
         L_\delta^2 \right)
\nonumber \\
\lefteqn{\Delta_{H}^s =  {727 \over 144} - {\pi^2 \over 2} 
       + {2 \over 9}\delta
+ \delta^2 \left( {1 \over 36} \pi^2 -{67 \over 2700} - {1 \over 30} L_\delta
        \right) }.
\nonumber \\
\ea

The results for the pseudo-scalar and axial-vector currents are the
same as for the scalar and vector currents because the presence of
massless fermion line permits one to cancel the Dirac $\gamma_5$
matrices.

It is interesting to compare our results with the numbers obtained in
\cite{Chetyrkin:2001je}.  We have plotted our results for the
independent color structures as functions of the velocity $v =
\delta/(2-\delta)$ using our results for $R_2^{v}$, including terms
$\order{\delta^5}$ which we do not display in \eq{eq:v}.  Comparing
these results with Fig.~4 of \cite{Chetyrkin:2001je}, we find very
good agreement for $v \le 0.6$ \cite{KGNumbers}; for larger values
of $v$ our truncated series is not accurate (as can be expected from
the very nature of expansion) and more terms in the expansion are
needed.  We have also verified the relations between the scalar and
vector correlators given in eqs. (32-33) in \cite{Chetyrkin:2001je}.

In \cite{Chetyrkin:2001je},  the values of the ${\cal O}(\delta^2 \ln^{0}
\delta)$ terms were estimated 
by fitting numerical solutions.  Our formulas provide analytic results
for these coefficients.  In the notation of eq.~(45) of
\cite{Chetyrkin:2001je}, we find
\ba
\tilde c_{FF} &=& \frac{1173}{128} + \frac{103\,{\pi }^2}{48} + 
  \frac{{\pi }^4}{90} -
\left( \frac{97}{16} +   \frac{5\,\pi^2}{6} \right)\ln 2
\nonumber \\
&&
+ \frac{9}{8}\ln^2 2 - \frac{\zeta_3}{2} \simeq 21.46,
\nonumber \\
\tilde c_{FA} &=& \frac{20057}{1152} + \frac{119\,{\pi }^2}{216} - 
  \frac{{\pi }^4}{90} 
-\left( \frac{141}{16} +
  \frac{19\,\pi^2}{36}\right) \ln 2 
\nonumber \\ &&
+ 
  \frac{11}{8} \ln^2 2 - 
  \frac{13\,\zeta_3}{2} \simeq 4.894,
\nonumber \\
\tilde c_{FL} &=&
- \frac{1849}{288}   - 
  \frac{23\,{\pi }^2}{108} + \left(\frac{13}{4} + 
  \frac{2\,{\pi }^2}{9}\right) \ln 2
\nonumber \\ &&
 - \frac{{\ln ^2 2}}{2} + 
  2\,\zeta_3
\simeq -2.585.
\ea
These results agree well with the estimates of \cite{Chetyrkin:2001je}
for the the abelian $\tilde c_{FF}=21(6)$ and the light quark $\tilde
c_{FL}=-2.3(7)$ contributions.  The non-abelian part found in
\cite{Chetyrkin:2001je}, $\tilde c_{FA}=1.2(4)$, differs by about 9
sigma from our result, 4.894.  Since the production threshold is the
most difficult place for the Pad\'e approximants
it is very likely \cite{KG} that the accuracy of this
particular result was overestimated in \cite{Chetyrkin:2001je}. 

The expansion around the heavy-light threshold presented here extends
the class of Feynman diagrams which can be evaluated analytically.
Our approach can be summarized as applying the HQET directly to
Feynman diagrams.  As an example application, we computed the
imaginary part of flavor off-diagonal current correlators to ${\cal
O}(\alpha_s^2)$ useful as an input for both $f_B$ determination from
QCD sum rules and also for matching of the lattice and continuum
currents.  Similar techniques can be used to study second order QCD
corrections to differential distributions in heavy to light
semileptonic decays.  Work on this is in progress.

We are grateful to G. P. Lepage for a conversation which stimulated
this study. This research was supported in part by the
Natural Sciences and Engineering Research Council of Canada and by the
DOE under grant number DE-AC03-76SF00515.


\begin{thebibliography}{18}
\expandafter\ifx\csname natexlab\endcsname\relax\def\natexlab#1{#1}\fi
\expandafter\ifx\csname bibnamefont\endcsname\relax
  \def\bibnamefont#1{#1}\fi
\expandafter\ifx\csname bibfnamefont\endcsname\relax
  \def\bibfnamefont#1{#1}\fi
\expandafter\ifx\csname citenamefont\endcsname\relax
  \def\citenamefont#1{#1}\fi
\expandafter\ifx\csname url\endcsname\relax
  \def\url#1{\texttt{#1}}\fi
\expandafter\ifx\csname urlprefix\endcsname\relax\def\urlprefix{URL }\fi
\providecommand{\bibinfo}[2]{#2}
\providecommand{\eprint}[2][]{\url{#2}}

\bibitem[{\citenamefont{Khodjamirian}(2001)}]{Khodjamirian:2001bj}
For a recent review  and references to original papers see 
\bibinfo{author}{\bibfnamefont{A.}~\bibnamefont{Khodjamirian}}
  (\bibinfo{year}{2001}), \eprint{hep-ph/0108205}.

\bibitem[{lep()}]{lepage}
\bibinfo{note}{We are indebted to G. P. Lepage for a discussion of this point.}

\bibitem[{\citenamefont{Chetyrkin}(1991)}]{Chetyrkin91}
\bibinfo{author}{\bibfnamefont{K.~G.} \bibnamefont{Chetyrkin}}
  (\bibinfo{year}{1991}), \bibinfo{note}{preprint MPI-Ph/PTh 13/91}.

\bibitem[{\citenamefont{Tkachev}(1994)}]{Tkachev:1994gz}
\bibinfo{author}{\bibfnamefont{F.~V.} \bibnamefont{Tkachev}},
  \bibinfo{journal}{Sov. J. Part. Nucl.} \textbf{\bibinfo{volume}{25}},
  \bibinfo{pages}{649} (\bibinfo{year}{1994}), \eprint{hep-ph/9701272}.

\bibitem[{\citenamefont{Smirnov}(1995)}]{Smirnov:1995tg}
\bibinfo{author}{\bibfnamefont{V.~A.} \bibnamefont{Smirnov}},
  \bibinfo{journal}{Mod. Phys. Lett.} \textbf{\bibinfo{volume}{A10}},
  \bibinfo{pages}{1485} (\bibinfo{year}{1995}), \eprint{hep-th/9412063}.

\bibitem[{\citenamefont{Smirnov}(1997)}]{Smirnov:1997gx}
\bibinfo{author}{\bibfnamefont{V.~A.} \bibnamefont{Smirnov}},
  \bibinfo{journal}{Phys. Lett.} \textbf{\bibinfo{volume}{B404}},
  \bibinfo{pages}{101} (\bibinfo{year}{1997}), \eprint{hep-ph/9703357}.

\bibitem[{\citenamefont{Czarnecki and Smirnov}(1997)}]{CzSmir}
\bibinfo{author}{\bibfnamefont{A.}~\bibnamefont{Czarnecki}} \bibnamefont{and}
  \bibinfo{author}{\bibfnamefont{V.}~\bibnamefont{Smirnov}},
  \bibinfo{journal}{Phys. Lett.} \textbf{\bibinfo{volume}{B394}},
  \bibinfo{pages}{211} (\bibinfo{year}{1997}), \eprint{hep-ph/9608407}.

\bibitem[{\citenamefont{Czarnecki and Melnikov}(2001)}]{Czarnecki:2000fv}
\bibinfo{author}{\bibfnamefont{A.}~\bibnamefont{Czarnecki}} \bibnamefont{and}
  \bibinfo{author}{\bibfnamefont{K.}~\bibnamefont{Melnikov}},
  \bibinfo{journal}{Phys. Rev. Lett.} \textbf{\bibinfo{volume}{87}},
  \bibinfo{pages}{013001} (\bibinfo{year}{2001}), \eprint{hep-ph/0012053}.

\bibitem[{\citenamefont{Beneke and Smirnov}(1998)}]{Beneke:1998zp}
\bibinfo{author}{\bibfnamefont{M.}~\bibnamefont{Beneke}} \bibnamefont{and}
  \bibinfo{author}{\bibfnamefont{V.~A.} \bibnamefont{Smirnov}},
  \bibinfo{journal}{Nucl. Phys.} \textbf{\bibinfo{volume}{B522}},
  \bibinfo{pages}{321} (\bibinfo{year}{1998}), \eprint{hep-ph/9711391}.

\bibitem[{\citenamefont{Chetyrkin and
  Steinhauser}(2001{\natexlab{a}})}]{Chetyrkin:2000mq}
\bibinfo{author}{\bibfnamefont{K.~G.} \bibnamefont{Chetyrkin}}
  \bibnamefont{and}
  \bibinfo{author}{\bibfnamefont{M.}~\bibnamefont{Steinhauser}},
  \bibinfo{journal}{Phys. Lett.} \textbf{\bibinfo{volume}{B502}},
  \bibinfo{pages}{104} (\bibinfo{year}{2001}{\natexlab{a}}),
  \eprint{hep-ph/0012002}.

\bibitem[{\citenamefont{Chetyrkin and
  Steinhauser}(2001{\natexlab{b}})}]{Chetyrkin:2001je}
\bibinfo{author}{\bibfnamefont{K.~G.} \bibnamefont{Chetyrkin}}
  \bibnamefont{and}
  \bibinfo{author}{\bibfnamefont{M.}~\bibnamefont{Steinhauser}},
  \bibinfo{journal}{Eur. Phys. J.} \textbf{\bibinfo{volume}{C21}},
  \bibinfo{pages}{319} (\bibinfo{year}{2001}{\natexlab{b}}),
  \eprint{hep-ph/0108017}.

\bibitem[{\citenamefont{Laporta and Remiddi}(1993)}]{Laporta:1993pa}
\bibinfo{author}{\bibfnamefont{S.}~\bibnamefont{Laporta}} \bibnamefont{and}
  \bibinfo{author}{\bibfnamefont{E.}~\bibnamefont{Remiddi}},
  \bibinfo{journal}{Phys. Lett.} \textbf{\bibinfo{volume}{B301}},
  \bibinfo{pages}{440} (\bibinfo{year}{1993}).

\bibitem[{\citenamefont{Melnikov and van Ritbergen}(2000)}]{Melnikov:2000zc}
\bibinfo{author}{\bibfnamefont{K.}~\bibnamefont{Melnikov}} \bibnamefont{and}
  \bibinfo{author}{\bibfnamefont{T.}~\bibnamefont{van Ritbergen}},
  \bibinfo{journal}{Nucl. Phys.} \textbf{\bibinfo{volume}{B591}},
  \bibinfo{pages}{515} (\bibinfo{year}{2000}), \eprint{hep-ph/0005131}.

\bibitem[{\citenamefont{Grozin}(2000)}]{Grozin:2000jv}
\bibinfo{author}{\bibfnamefont{A.~G.} \bibnamefont{Grozin}},
  \bibinfo{journal}{JHEP} \textbf{\bibinfo{volume}{03}}, \bibinfo{pages}{013}
  (\bibinfo{year}{2000}), \eprint{hep-ph/0002266}.

\bibitem[{\citenamefont{Grozin}(2001)}]{Grozin:2001fw}
\bibinfo{author}{\bibfnamefont{A.~G.} \bibnamefont{Grozin}}
  (\bibinfo{year}{2001}), \eprint{hep-ph/0107248}.

\bibitem[{\citenamefont{Vermaseren}()}]{form3}
\bibinfo{author}{\bibfnamefont{J.~A.~M.} \bibnamefont{Vermaseren}},
  \bibinfo{note}{math-ph/0010025}.

\bibitem[{\citenamefont{Beneke and Braun}(1994)}]{Beneke:1994sw}
\bibinfo{author}{\bibfnamefont{M.}~\bibnamefont{Beneke}} \bibnamefont{and}
  \bibinfo{author}{\bibfnamefont{V.~M.} \bibnamefont{Braun}},
  \bibinfo{journal}{Nucl. Phys.} \textbf{\bibinfo{volume}{B426}},
  \bibinfo{pages}{301} (\bibinfo{year}{1994}), \eprint{hep-ph/9402364}.

\bibitem[{\citenamefont{Chetyrkin and Tkachev}(1981)}]{che81}
\bibinfo{author}{\bibfnamefont{K.~G.} \bibnamefont{Chetyrkin}}
  \bibnamefont{and} \bibinfo{author}{\bibfnamefont{F.~V.}
  \bibnamefont{Tkachev}}, \bibinfo{journal}{Nucl. Phys.}
  \textbf{\bibinfo{volume}{B192}}, \bibinfo{pages}{159}
  (\bibinfo{year}{1981}). 



\bibitem{KGNumbers} We are grateful to K. G. Chetyrkin for providing
sample numerical values for comparisons. 

\bibitem{KG} K. G. Chetyrkin, private communication.

\end{thebibliography}

\end{document}